\documentclass[aps,prb]{revtex4}
\usepackage{amsmath,amssymb,graphicx}
\begin{document}
\narrowtext

\title{Excitons in narrow-gap carbon nanotubes}

\author{R. R. Hartmann}
\affiliation{School of Physics, University of Exeter, Stocker Road,
Exeter EX4 4QL, United Kingdom}

\author{I. A. Shelykh}
\affiliation{Science Institute, University of Iceland, Dunhagi 3, IS-107, Reykjavik, Iceland}
\affiliation{International Institute of Physics, Av. Odilon Gomes de Lima, 1722, Capim Macio, CEP: 59078-400, Natal - RN, Brazil}

\author{M. E. Portnoi}
\email[Corresponding author: ]{M.E.Portnoi@exeter.ac.uk}
\affiliation{School of Physics, University of Exeter, Stocker Road,
Exeter EX4 4QL, United Kingdom}
\affiliation{International Institute of Physics, Av. Odilon Gomes de Lima, 1722, Capim Macio, CEP: 59078-400, Natal - RN, Brazil}

\begin{abstract}
We calculate the exciton binding energy in single-walled carbon nanotubes with
narrow band gaps, accounting for the quasi-relativistic dispersion
of electrons and holes. Exact analytical solutions of the
quantum relativistic two-body problem are obtain for several limiting cases.
We show that the binding energy scales with the band gap, and conclude on
the basis of the data available for semiconductor nanotubes that there
is no transition to an excitonic insulator in quasi-metallic nanotubes and
that their THz applications are feasible.
\end{abstract}
\date{August 15, 2011}

\pacs{78.67.Ch, 73.22.Pr, 73.63.Fg, 71.35.-y}
\maketitle

\section{Introduction}

In the 1940s it was predicted by P. R. Wallace \cite{Wallace1947}
that due to its honeycomb lattice, graphene, a single monolayer of
carbon, should exhibit unusual semimetallic behavior. The gap
between the valence and conductance band is exactly zero, and the
low-energy excitations are massless chiral Dirac fermions.
\cite{NovoselovNature,CastroNetoRMP} On the other hand, the
electronic band structure of other carbon-based materials shows
substantial differences from the band structure of graphene. Among
them are fullerenes, which can be considered as zero-dimensional
carbon molecules with discrete energy spectra, and carbon nanotubes,
which are obtained by rolling graphene along a given direction and
reconnecting the carbon bonds.\cite{Charlier2007}

The energy spectrum of a single-walled carbon nanotube is determined by the way it
is rolled \cite{Saito} and is characterized by two integers $(n,m)$
($0\leq m \leq n$) denoting the relative position,
$\textbf{C}_h=n\textbf{a}_1+m\textbf{a}_2$, of the pair of atoms on
a graphene strip which coincide when the strip is rolled into a tube
($\textbf{a}_1$ and $\textbf{a}_2$ are the unit vectors of the
hexagonal lattice). For most combinations of $n$ and $m$ the energy
spectrum of the nanotube is characterized by the gap, whose value is
comparable to those in semiconductor materials. However, for a third
of $n$ and $m$ combinations, namely when $n-m=3p,\; p=0, 1, 2,...$
the value of the gap is drastically reduced and lies in terahertz
frequency range. Moreover, for $m=n$ the gap vanishes in zero
magnetic field, and opens only after the application of a magnetic
field parallel to the nanotube axis.
\cite{Charlier2007,Saito,Ando93}

Optical properties of carbon nanotubes have been investigated by
many authors.\cite{Bachilo,Hartschuh,Wang,WangScience05,Berciaud}
It was shown that the excitonic effect plays an important role and
that the properties of the excitonic resonance can be modulated by
applying external fields.\cite{Shaver,Shaver1,Mohite} However,
excitons were theoretically studied mostly for semiconductor carbon
nanotubes with sufficiently large gaps.
\cite{Pedersen,PerebeinosPRL,AndoTransverse,AndoReview09}
For metallic nanotubes mostly the excitons associated with high branches
of the nanotube spectrum separated by an energy of about $2\;$eV were
considered.\cite{metallic,metallic2}  On the other hand, the
analysis of the long-wavelength properties of narrow-band nanotubes
is also of high interest, as there is a growing number of proposals
using carbon nanotubes of this type for THz applications, including
several schemes put forward by the authors of the present work.
\cite{PortnoiNano,PortnoiSM,PortnoiJMPB}
The only work on excitons in narrow-gap nanotubes,
in which the stability of excitons in metallic carbon nanotubes subjected
to Aharonov-Bohm magnetic flux was considered is Ref.~\onlinecite{Ando1997},
where it was claimed that even in this case the exciton
binding energy never exceeds the gap.
The main conclusions of Ref.~\onlinecite{Ando1997} are based on numerical calculations
in the ${\textbf{k}}\cdot{\textbf{p}}$ scheme.
In our present work we give further consideration to this interesting case, using a
noticeably different semi-analytical approach based on a Dirac-like matrix equation.

At first glance, narrow-gap carbon nanotubes fit ideally the excitonic insulator
picture \cite{Jerome} predicted for the case when the exciton binding
energy exceeds the band gap. However, the energy spectrum of a
narrow-gap carbon nanotube is quasi-relativistic with the effective mass of the
charge carriers proportional to the gap leading to a drastic reduction
of the exciton binding energy with reducing the gap.  In this paper
we consider excitons formed by relativistic quasi-one-dimensional
electrons and holes interacting via a model yet realistic potential.
We show that the binding energy scales with the band gap and
conclude on the basis of the data available for semiconductor
nanotubes that there is no transition to an excitonic insulator in
quasi-metallic carbon nanotubes and that their THz applications are feasible.

\section{Formalism and interaction potential}

The Hamiltonian for a single free electron in a narrow-gap carbon nanotube
can be written (see Appendix \ref{App:magfield}) as
\begin{equation}
\widehat{H}_0=\hbar v_{\textrm{\scriptsize{F}}}\left(\begin{array}{cc}
0& b\widehat{q}-i\Delta\\
b\widehat{q}+i\Delta & 0\end{array}\right),
\label{H0}
\end{equation}
where $\hat{q}=-i\frac{\partial}{\partial x}$ is the operator of the wavevector along the
nanotube axis and we use the basis
$\left|\psi_A\right\rangle,\left|\psi_B\right\rangle $ with indices
$A,B$ corresponding to the carbon atoms of two different sublattices
in the honeycomb lattice. Here $v_{\textrm{\scriptsize{F}}}$ is the
Fermi velocity in graphene, connected to the tight-binding matrix
element of electron hopping $|t| \approx 3\;$eV and the graphene
lattice constant $a$ by $\hbar
v_{\textrm{\scriptsize{F}}}=\frac{\sqrt{3}}{2}|t| a$, where
$a=|\textbf{a}_1|=|\textbf{a}_2|=2.46\;$\r{A}.\cite{Saito} For the
$(n,n)$ armchair nanotube the value of the band gap $2\hbar
v_{\textrm{\scriptsize{F}}}|\Delta|$ is determined by the external
magnetic field, $\Delta=\frac{2}{a
\sqrt{3}}\sin\left(\frac{\pi\Phi}{n\Phi_0}\right)$, where $\Phi$ is
the magnetic flux through the nanotube cross section, $\Phi_0=ch/e$
is the flux quantum and
$b=\sqrt{\frac{4}{3}-\frac{1}{3}\cos^2\left(\frac{\pi\Phi}{n\Phi_0}\right)}$.
Since for experimentally accessible magnetic fields
$\Phi/\Phi_0\ll1$, in our further consideration we set $b=1$. A
similar Hamiltonian can be written for a narrow-gap carbon nanotube
with a gap opened by curvature \cite{Kane} or for certain types of
graphene nanoribbons \cite{Fertig}. The diagonalization of
equation~(\ref{H0}) gives a quasi-relativistic dispersion, $E=\pm\hbar
v_{\textrm{\scriptsize{F}}}\sqrt{\Delta^2+q^2}$. To go from the case
of the electrons to the case of the holes,
$|\psi_i^e\rangle\rightarrow|\psi_i^h\rangle$ one should use the
substitution $\widehat{q}\rightarrow -\widehat{q},~\Delta\rightarrow
-\Delta$. For a pair of interacting electron and hole the total
Hamiltonian can be written in the form of a $4\times4$ matrix, and
the stationary Schr\"{o}dinger equation for determining the binding
energy  written in the basis
$|\Psi_{ij}\rangle=|\psi_i^e\rangle|\psi_j^h\rangle$ reads
\begin{widetext}
\begin{equation}
\hbar v_{\textrm{\scriptsize{F}}}\left(\begin{array}{cccc}
0 & \widehat{q}_{e}-i\Delta & -\widehat{q}_{h}+i\Delta & 0\\
\widehat{q}_{e}+i\Delta & 0 & 0 & -\widehat{q}_{h}+i\Delta\\
-\widehat{q}_{h}-i\Delta & 0 & 0 & \widehat{q}_{e}-i\Delta\\
0 & -\widehat{q}_{h}-i\Delta & \widehat{q}_{e}+i\Delta & 0\end{array}\right)\left(\begin{array}{c}
\Psi_{AA}\\
\Psi_{BA}\\
\Psi_{AB}\\
\Psi_{BB}\end{array}\right)=\left[E-V(x_e-x_h)\right]\left(\begin{array}{c}
\Psi_{AA}\\
\Psi_{BA}\\
\Psi_{AB}\\
\Psi_{BB}\end{array}\right),
\label{SchrodingerPair}
\end{equation}
\end{widetext}
where the indices $e$ and $h$ correspond to the electrons and holes, and
$\widehat{q}_{e,h}=-i\frac{\partial}{\partial x_{e,h}}$. In the
absence of interaction and band-filling effects this Hamiltonian
yields four energy eigenvalues corresponding to a pair of
non-interacting quasi-particles:  $E=\hbar
v_{\textrm{\scriptsize{F}}}(\pm \sqrt{\Delta^2+q_{e}^2} \pm
\sqrt{\Delta^2+q_{h}^2})$; only the solution with positive signs should
be taken if we consider a system containing a single electron and a single hole.
As the potential of electron-hole interaction
$V(x_e-x_h)$ depends only on the distance between the electron and
hole, it is convenient to use the center of mass and relative motion
coordinates,
$X=(x_e+x_h)/2,x=x_e-x_h,~\widehat{q}_{e}=\widehat{K}/2+\widehat{k},~\widehat{q}_{h}=\widehat{K}/2-\widehat{k}$,
and represent the exciton wavefunctions as
$\Psi_{ij}(X,x)=e^{iKX}\phi_{ij}(x)$, which permits the substitution
of the operator $\widehat{K}$ by the number $K$, having the physical
meaning of the wavevector of the exciton as a whole. Considering the
case of $K=0$ corresponding to a static exciton, the equation for
the wavefunction of relative motion reads
\begin{widetext}
\begin{equation}
\left(\begin{array}{cccc}
0 & \widehat{k}-i\Delta & \widehat{k}+i\Delta & 0\\
\widehat{k}+i\Delta & 0 & 0 & \widehat{k}+i\Delta\\
\widehat{k}-i\Delta & 0 & 0 & \widehat{k}-i\Delta\\
0 & \widehat{k}-i\Delta & \widehat{k}+i\Delta & 0\end{array}\right)\left(\begin{array}{c}
\phi_{AA}\\
\phi_{BA}\\
\phi_{AB}\\
\phi_{BB}\end{array}\right)=\left[\varepsilon-\tilde{V}(x)\right]\left(\begin{array}{c}
\phi_{AA}\\
\phi_{BA}\\
\phi_{AB}\\
\phi_{BB}\end{array}\right),
\label{SchrodingerExciton}
\end{equation}
\end{widetext}
where $\varepsilon=E/\hbar v_{\textrm{\scriptsize{F}}},~\tilde{V}(x)=V(x)/\hbar v_F$ and $\widehat{k}=-i\partial/\partial x$.
Eq.~(\ref{SchrodingerExciton}) represents a system of first order differential equations, which can be reduced (see Appendix \ref{App:E0}) to a single second order equation for $\phi_{AA}$:
\begin{widetext}
\begin{equation}
\frac{d^{2}\phi_{AA}}{dx^{2}}+\frac{1}{\varepsilon-\tilde{V}(x)}\,\frac{d\tilde{V}(x)}{dx}\,\frac{d\phi_{AA}}{dx}
+\left[\left(\frac{\varepsilon-\tilde{V}(x)}{2}\right)^{2}-\Delta^{2}\right]\phi_{AA}=0
\label{PsiAA}
\end{equation}
\end{widetext}
Before solving Eq.~(\ref{PsiAA}), we need to specify the interaction
potential. It needs to possess the following properties: first, it
should remain finite as $x\rightarrow 0$ and for small $x$ scale as
$V(x)\approx -e^2/\left(\epsilon \sqrt{x^2+d^2}\right)$, where
$\epsilon$ is the effective dielectric constant and $d$ is the
short-range cut-off parameter, which is of the order of the nanotube
diameter; second, for large $x$ it should decay exponentially due to
the effects of screening necessarily present in nanotubes of the
metallic type \cite{metallic} - it is apparent that the size of the
exciton should not exceed the mean separation between
quasi-particles. The convenient choice of the potential is
\begin{equation}
\tilde{V}(x)=-\frac{\alpha}{\cosh(\beta x)}
\label{HamiltonianCosh}
\end{equation}
where $\alpha = e^2/(\epsilon \hbar v_{\textrm{\scriptsize{F}}} d)$
and the effective length, $1/\beta$, defines the spatial extension
of the interaction. The additional advantage of the potential given
by Eq.~(\ref{HamiltonianCosh}) is that it allows one to obtain some
analytical results in the context of graphene physics as was shown
in Ref.~\onlinecite{Richard}.

\section{Results and Discussion}

\begin{figure}[tbp]
\includegraphics[width=7.0cm]{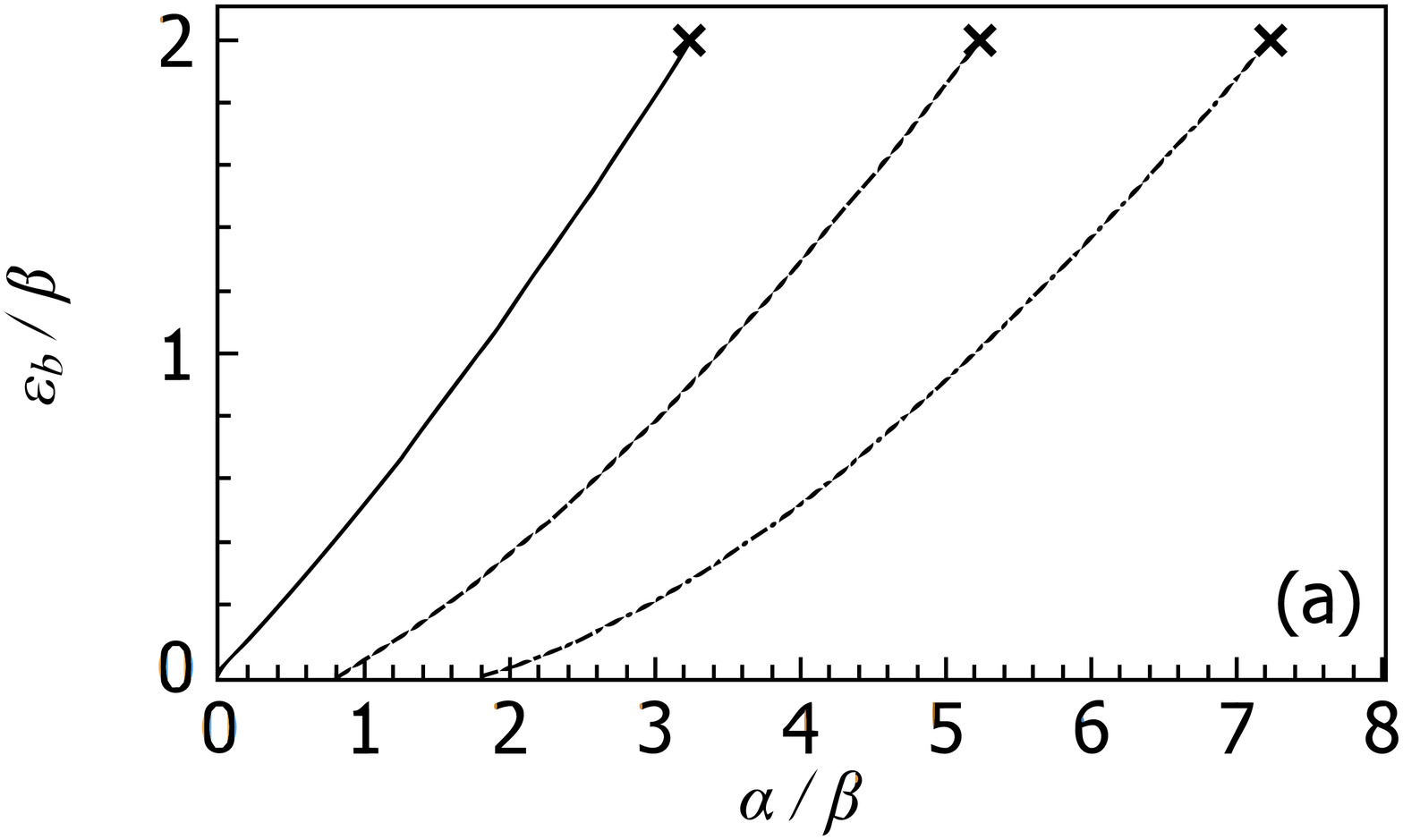}
\includegraphics[width=7.3cm]{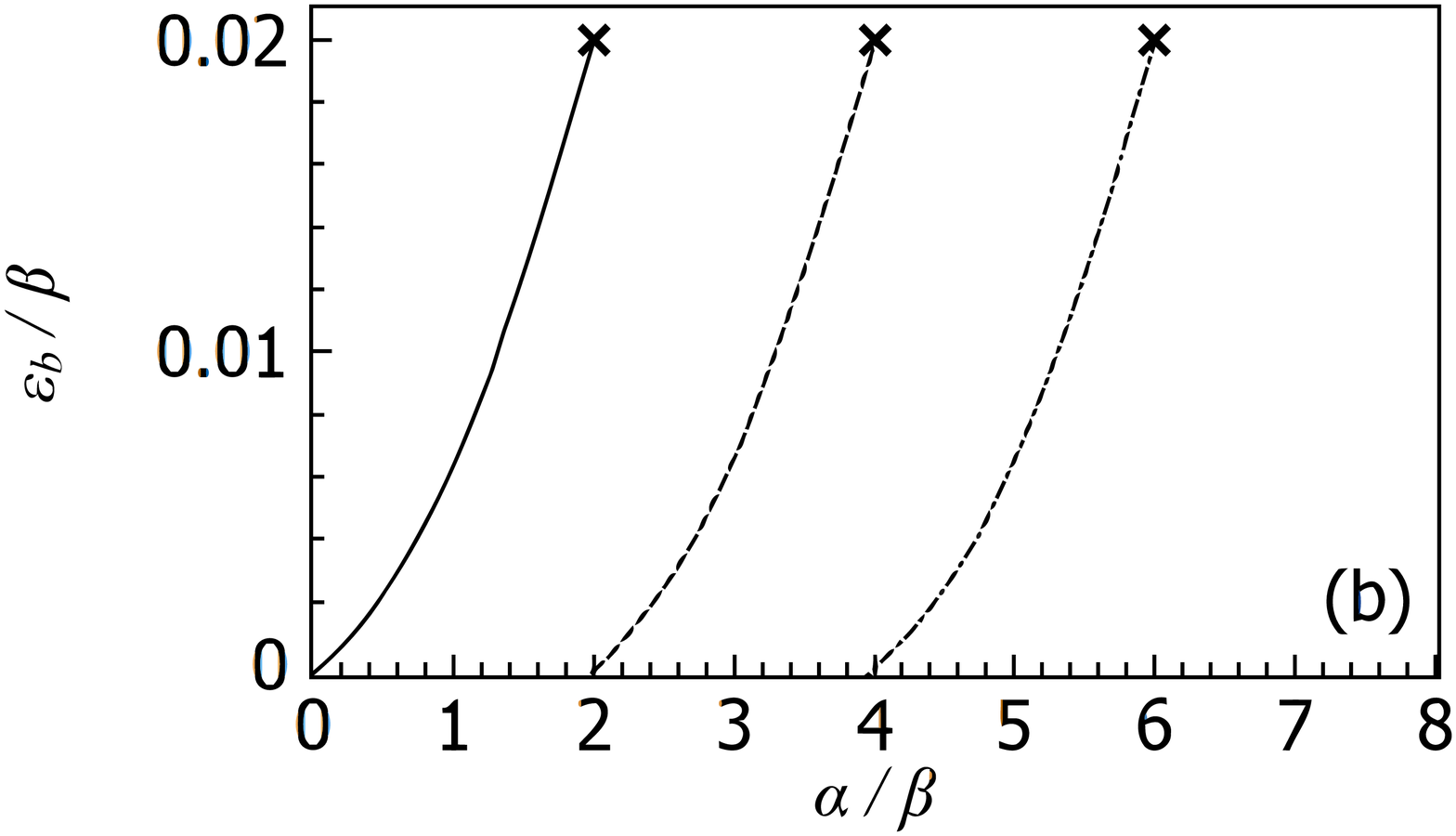}
\caption{The dependence of the exciton binding energy $\varepsilon_b/\beta$ on the interaction strength: (a) - for a semiconductor nanotube with $\Delta/\beta=1$; (b) - for a quasi-metallic nanotube with $\Delta/\beta=0.01$. The different lines correspond to different excitonic states. The crosses indicate the exact analytic results for zero-energy states ($\varepsilon_b=\varepsilon_g$).}
\label{Exciton_alpha}
\end{figure}

\begin{figure}
\includegraphics[width=12cm]{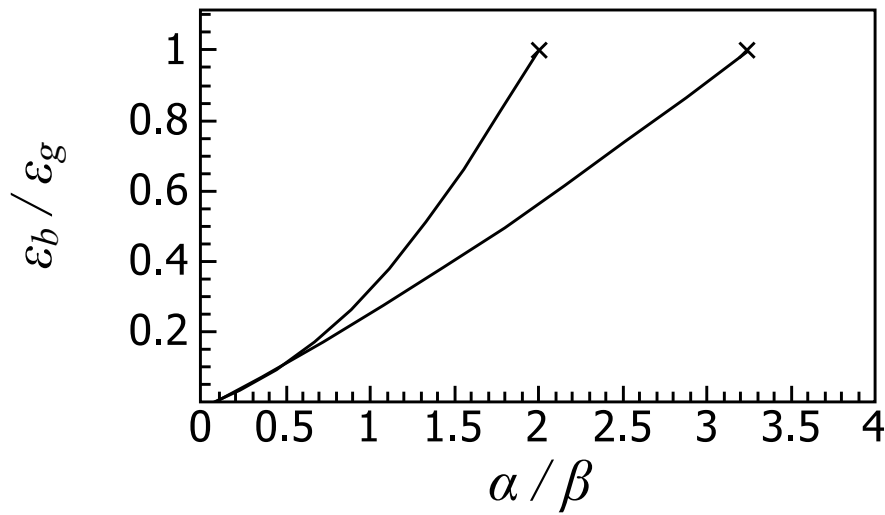}
\caption{The dependence of the ratio of the exciton binding energy to the nanotube band gap
on the interaction strength for a quasi-metallic carbon nanotube with $\Delta/\beta=0.01$
(upper curve) and a semiconductor nanotube with $\Delta/\beta=1$. Only s-states are shown.
The crosses correspond to the exact analytic results when the exciton binding energy is equal to the band gap.}
\label{Exciton_Delta}
\end{figure}

The results of the numerical solution are shown in
Fig.~\ref{Exciton_alpha} and Fig.~\ref{Exciton_Delta}. Figures
\ref{Exciton_alpha}(a) and \ref{Exciton_alpha}(b) show the
dependence of the binding energy $\varepsilon_b$ calculated as an
absolute value of the difference between the eigenenergy of
Eq.~(\ref{PsiAA}) and the energy of the pair of non-interacting
electron and hole, $\varepsilon_b=|\varepsilon-2\Delta|$, measured
in the units of $\beta$ versus the effective strength of the
interaction $\alpha/\beta$ for two different values of
$\Delta/\beta=1$ and $\Delta/\beta=0.01$ corresponding to the cases
of the semiconductor and narrow-gap quasi-metallic nanotube
respectively. In both cases, for small values of $\alpha/\beta$
there is only one bound s-type state, whose binding energy increases
with the increase of the interaction strength, until it reaches the
value $\varepsilon_b=\varepsilon_g=2\Delta$ after which it goes to
the continuum of states with negative energies and thus unbinds. On
the other hand, the increase of $\alpha/\beta$ leads to the
appearance of higher order solutions, corresponding to p, d,
etc. excitons. This is an adequate picture for a single
impurity-like exciton. However, if the interaction between an
electron and a hole created across the gap separating the ground and
excited states of the system exceeds the band gap in a many-body
system, the ground state should be redefined and the many-body
effects should govern the value of the gap. To find out whether the
gap in a narrow-gap nanotube is mostly governed by single-electron
effects such as an external magnetic field and curvature or by
many-body effects, the value of the effective interaction strength
$\alpha/\beta$ should be estimated from the known data for excitons
in semiconductor nanotubes.

Note also, that by reducing Eq.~(\ref{PsiAA}) for $\varepsilon=0$ to
the hypergeometric equation (see Appendix \ref{App:E0}) one can
find the analytical expression for the values of the parameter
$\alpha/\beta$ when the exciton binding energy is exactly equal to
the bandgap:
\begin{equation}
\frac{\alpha}{\beta}=1+2n+\sqrt{1+4\frac{\Delta^2}{\beta^2}},
\label{critical}
\end{equation}
where $n=0, 1, 2,...\;$. These exact values are shown by the crosses in Fig.~\ref{Exciton_alpha}.

It is convenient to express the dimensionless interaction strength $\alpha/\beta$ as
\begin{equation}
\frac{\alpha}{\beta}=\frac{e^2}{\epsilon d \hbar v_{\textrm{\scriptsize{F}}}\beta}=\frac{c}{v_{\textrm{\scriptsize{F}}}}
\frac{e^2}{\hbar c}
\frac{1}{\epsilon \beta d} \approx \frac{300}{137}\frac{1}{\epsilon \beta d}.
\label{strength}
\end{equation}
When the interaction strength is written in the above form, one can
immediately see the direct relation of the strongly-bound exciton
problem to the long-standing problem of a supercritical charge (the
nuclear charge with $Z>137$) and atomic collapse in relativistic
quantum mechanics \cite{Zeldovich71}. This problem has recently been
revisited and reformulated for graphene \cite{ShytovSSC09} where the
effective impurity charge is increased by a factor of
$c/v_{\textrm{\scriptsize{F}}}\approx 300$, which is also present in
the numerator of  Eq.~(\ref{strength}). It follows from
Eq.~(\ref{critical}) that for an exciton in a narrow-gap carbon
nanotube ($\Delta \ll \beta$) the effective supercritical charge
corresponding to $\varepsilon_b > \varepsilon_g$ can be achieved for
$\alpha/\beta>2$. In principle, parameters $\epsilon$ and $\beta$
can be controlled by submerging nanotubes in a solution or changing
the number of electrons and holes by injection, optical excitation or
varying temperature. However, one should not expect the value of
$1/(\epsilon \beta d)$ to be higher for a quasi-metallic carbon
nanotube than for a semiconductor nanotube of similar diameter.

Fig.~\ref{Exciton_Delta} shows the dependence of the ratio of the
binding energy of the 1s exciton to the value of the gap on the
interaction strength for the cases of the semiconductor and narrow
gap nanotubes. For very small interaction strengths the ratio
$\varepsilon_b/\varepsilon_g$ is a universal quadratic function of
$\alpha/\beta$ (see Appendix \ref{App:weak}).
The two curves remain practically indistinguishable from one another with increasing
$\alpha/\beta$ up to $\alpha/\beta\approx0.6$.
A further increase of the interaction strength or range leads to the breakdown of
universality for finite range potentials.

It should be noted, that for a zero-range attractive potential between the electron
and hole \cite{Roemer09} this universality holds for the whole range of
interaction strength (see Appendix \ref{App:deltafunction}). However, for the
smooth interaction potential considered here the universality breaks
for sufficiently large interaction strength. The difference between
the case of semiconductor and quasi-metallic nanotubes is not as large as one could expected.
Namely, the value of $\alpha/\beta$ for which the 1s-exciton binding energy is exactly equal
to the band gap is about 3.2 for the case of a semiconductor
nanotube with $\Delta/\beta=1$ and about 2 for a quasi-metallic
nanotube with $\Delta/\beta=0.01$. Also, the variational
calculations \cite{Pedersen} of the binding energy in semiconductor
nanotubes supported by the experimental data gives a value for the
exciton binding energy of approximately 30 percent of the bandgap
which according to Fig.~\ref{Exciton_Delta} corresponds to the
values of $\alpha/\beta\approx 1$.
Using the experimental data for the exciton binding energy for the excited
subbands of metallic tubes \cite{Shaver1} provides even smaller values for $\alpha/\beta$.
These estimations ensure that for a quasi-metallic nanotube the 1s exciton bound state lies
within the gap.
The hypothetical situation $\alpha/\beta> 2$ corresponds to the case of the so-called
``excitonic insulator'', \cite{Jerome} where the account of many body effects becomes
crucially important and goes beyond the scope of our consideration,
which is essentially a two-particle approach. Certain many-body
aspects of the physics of excitons in narrow gap nanotubes and their
relation to the Luttinger liquid model have been recently studied
using a numerical renormalization group.\cite{Konik10}

\begin{figure}[h]
\includegraphics[width=12cm]{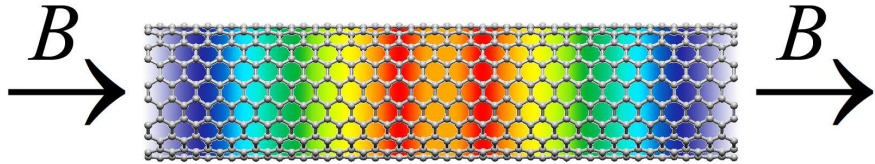}
\caption{(Color on-line) The density of the 1s-exciton for a (10,10) carbon nanotube with a magnetic field induced gap of $10\;$meV ($2.5\;$THz) corresponding to a magnetic field of $15\;$T along the nanotube axis. The density represents the probability of finding the electron and hole comprising the exciton at the indicated relative separation. Red and blue colors correspond to the highest and lowest values of density, respectively.}
\label{Wavefunction}
\end{figure}

Fig. \ref{Wavefunction} shows the square modulus of the excitonic wavefunction in real
space for a quasi-metallic nanotube. One sees that for the 1s state the density in the
center of mass has a local minimum, which differs strikingly from the result obtained
earlier for semiconductor nanotubes in the effective-mass approximation,\cite{WangScience05}
in which the probability density in the exciton center of mass has a maximum.
This is related to the complex matrix structure of the Hamiltonian (\ref{H0})
resulting in the multi-component structure of the eigenfunctions.
A similar dip in the ground-state density was previously reported for graphene-based
waveguides.\cite{Richard}

It should be noted that taking into account the valley and spin
quantum numbers increases the number of different types of excitons
associated with a given carbon nanotube spectrum branch to
sixteen.\cite{AndoReview09} Their consideration, however, can be
done along the lines described above, the only difference being the
modified interaction strength $\alpha/\beta$ for each type of exciton.
The exciton with the highest binding energy in a
semiconductor nanotube is known to be optically inactive (dark).
The difference between the energy of the dark and bright excitons is
proportional to the exciton binding energy and exceeds
$k_{\textrm{\scriptsize{B}}}T$ at room temperature causing
a significant suppression in the optical emission from semiconductor
nanotubes. As it is shown above, for narrow-gap nanotubes the
binding energy is drastically reduced and there should be no
noticeable difference in the population of dark and bright excitonic
states at any experimentally attainable temperatures. At room
temperature, all dark and bright excitons in narrow-gap nanotubes
should be fully ionized and the direct inter-band transitions
\cite{PortnoiNano,PortnoiSM,PortnoiJMPB} govern the emission in the
terahertz range.

\section{Conclusions}

In conclusion, we considered the formation of the exciton in narrow
gap carbon nanotubes characterized by the quasi-relativistic spectrum
of free particles. We show that the exciton binding energy scales
with the band gap and vanishes with decreasing the gap even for strong
electron-hole attraction. Therefore, excitonic effects including
strongly-bound dark excitons, which explain the poor electroluminescent
properties of semiconducting nanotubes, should not dominate for narrow-gap
carbon nanotubes. This opens the possibility of using quasi-metallic
carbon nanotubes for various terahertz applications.

\begin{acknowledgments}
This work was supported by FP7 ITN Spinoptronics (Grant No. FP7-237252) and
FP7 IRSES projects SPINMET (Grant No. FP7-246784), TerACaN (Grant No. FP7-230778),
and ROBOCON (Grant No. FP7-230832).
I.A.S. acknowledges the support from Rannis ``Center of excellence in polaritonics''.
\end{acknowledgments}


\appendix
\section{Armchair carbon nanotube in a magnetic field}
\label{App:magfield}

Graphene's effective matrix Hamiltonian following the notations of
Ref.~\onlinecite{Saito} is given by
\[
H=t\left(\begin{array}{cc}
0 & f_{k}\\
f_{k}^{\star} & 0\end{array}\right)\]
and the corresponding eigenvalues are\[
E=\pm\left|t\right|\sqrt{\left|f_{k}\right|^{2}},\]
where \begin{equation}
f_{k}=\exp\left(i\frac{a}{\sqrt{3}}k_{x}\right)
+2\exp\left(-i\frac{a}{2\sqrt{3}}k_{x}\right)\cos\left(k_{y}
\frac{a}{2}\right).\label{eq:f}
\end{equation}

For an $\left(n,n\right)$ armchair carbon nanotube, $k_{x}$
is quantized in the following manner
\[k_{x}=\frac{2\pi}{a\sqrt{3}}\frac{l}{n},\]
where $l$ is an integer. Defining $k_{T}$ as the projection of the
wavevector along the nanotube axis Eq.~(\ref{eq:f}) can be expressed as
\begin{equation}
f_{k}=\exp\left(i\frac{2\pi}{3}\frac{l}{n}\right)
+2\exp\left(-i\frac{\pi}{3}\frac{l}{n}\right)\cos\left(\frac{k_{T}a}{2}\right).
\label{eq:f_l}
\end{equation}

In the presence of a magnetic field, $l\rightarrow l+\mathcal{F}$
(here $\mathcal{F}=\Phi/\Phi_{0}$) and the armchair carbon nanotube
energy spectrum becomes
\begin{equation}
E=\pm\left|t\right|\sqrt{1+4\cos\left(\pi\frac{l+\mathcal{F}}{n}\right)
\cos\left(\frac{k_{T}a}{2}\right)+4\cos^{2}\left(\frac{k_{T}a}{2}\right)}.
\end{equation}

We are interested solely in the lowest conduction and highest valence
subbands, which correspond to $l=n$. In this instance the subbands
energy spectra are
\begin{equation}
E=\pm\left|t\right|\sqrt{1-4\cos\left(\pi\frac{\mathcal{F}}{n}\right)
\cos\left(\frac{k_{T}a}{2}\right)+4\cos^{2}\left(\frac{k_{T}a}{2}\right)},
\label{eq:arm_chair_spectrum_mag}
\end{equation}
where the $+\,\left(-\right)$ sign denotes the lowest conduction
(highest valence) subband. The introduction of a magnetic field along
the nanotube axis shifts the minimum in the spectrum away from
$k_{T}=2\pi/3a$, the electrons {}``acquire mass'' and a band gap is opened.
The new minimum, $k_{T}^{\mathrm{min}}$, is obtained by differentiating
Eq.(\ref{eq:arm_chair_spectrum_mag}) and equating it to zero, this
yields:
\begin{equation}
\frac{1}{2}\cos\left(\pi\frac{\mathcal{F}}{n}\right)=\cos\left(k_{T}^{\mathrm{min}}\frac{a}{2}\right).
\label{eq:def_of_min}
\end{equation}

Since we are interested in particle behavior in the vicinity of $k_{T}^{\mathrm{min}}$,
it is natural to re-express the electron energy spectrum in terms
of $q_{T}$, defined as the momentum measured relative to $k_{T}^{\mathrm{min}}$
i.e. $k_{T}=k_{T}^{\mathrm{min}}+q_{T}$, thus Eq. (\ref{eq:f_l})
can be expressed by
\begin{widetext}
\[
f_{k}=\exp\left[-i\frac{\pi}{3}\left(1+\frac{\mathcal{F}}{n}\right)\right]
\left\{ -\exp\left(i\pi\frac{\mathcal{F}}{n}\right)+
2\cos\left[\left(k_{T}^{min}+q_{T}\right)\frac{a}{2}\right]\right\}.
\]
Using the identity Eq.~(\ref{eq:def_of_min}) $f_{k}$ becomes:
\begin{equation}
f_{k}=\exp\left[-i\frac{\pi}{3}\left(1+\frac{\mathcal{F}}{n}\right)\right]
\left[\cos\left(\pi\frac{\mathcal{F}}{n}\right)\cos\left(\frac{aq_{T}}{2}\right)
-2\sqrt{1-\frac{1}{4}\cos^{2}\left(\pi\frac{\mathcal{F}}{n}\right)}
\sin\left(\frac{aq_{T}}{2}\right)-\exp\left(i\pi\frac{\mathcal{F}}{n}\right)\right].
\label{eq:f_1}
\end{equation}
Expanding Eq.~(\ref{eq:f_1}) in terms of $q_{T}$ and retaining first
order terms only yields
\begin{equation}
f_{k}=\exp\left[-i\frac{\pi}{3}\left(1+\frac{\mathcal{F}}{n}\right)\right]
\left(-i\sin\left(\pi\frac{\mathcal{F}}{n}\right)-
\sqrt{1-\frac{1}{4}\cos^{2}\left(\pi\frac{\mathcal{F}}{n}\right)}aq_{T}\right).
\end{equation}
The effective matrix Hamiltonian can therefore be written as:
\begin{equation}
\hat{H}_{0}=t\left(\begin{array}{cc}
0 & e^{-i\theta}
\left(-i\sin\left(\pi\frac{\mathcal{F}}{n}\right)-
\sqrt{1-\frac{1}{4}\cos^{2}\left(\pi\frac{\mathcal{F}}{n}\right)}a\hat{q}\right)\\
e^{i\theta}\left(i\sin\left(\pi\frac{\mathcal{F}}{n}\right)-
\sqrt{1-\frac{1}{4}\cos^{2}\left(\pi\frac{\mathcal{F}}{n}\right)}a\hat{q}\right) & 0\end{array}\right),
\end{equation}
where $\theta=\frac{\pi}{3}\left(1+\frac{\mathcal{F}}{n}\right)$ and $\hat{q}=-i\frac{\partial}{\partial y}$.
This $2\times 2$ Hamiltonian acts on a two-component Dirac wavefunction with components $\chi_{1}$ and $\chi_{2}$ associated with the $A$ and $B$ sublattices of graphene respectively. By changing the basis
wavefunctions from $\chi_{1}$ and $\chi_{2}$ to $\psi_{A}=-\chi_{1}$ and
$\psi_{B}=-\exp(-i\theta)\chi_{2}$, and changing the variable $y\rightarrow-x$, so that the $x$-coordinate
is now along the nanotube axis, the effective matrix Hamiltonian can be expressed as
\begin{equation}
\hat{H}_{0}=\left|t\right|\left(\begin{array}{cc}
0 & \sqrt{1-\frac{1}{4}\cos^{2}\left(\pi\frac{\mathcal{F}}{n}\right)}a\hat{q}-i\sin\left(\pi\frac{\mathcal{F}}{n}\right)\\
\sqrt{1-\frac{1}{4}\cos^{2}\left(\pi\frac{\mathcal{F}}{n}\right)}a\hat{q}+i\sin\left(\pi\frac{\mathcal{F}}{n}\right) & 0\end{array}\right),
\end{equation}
which acts on the two-component Dirac wavefunction $(\psi_{A}, \psi_{B})^{\textrm{T}}$.
For brevity let $b=\sqrt{\frac{4}{3}-\frac{1}{3}\cos^{2}\left(\frac{\pi}{n}{\mathcal{F}}\right)}$
and $\Delta=\frac{2}{a\sqrt{3}}\sin\left(\frac{\pi}{n}{\mathcal{F}}\right)$.
Thus,
\begin{equation}
\hat{H}_{0}=\hbar v_{\mathrm{F}}\left(\begin{array}{cc}
0 & b\hat{q}-i\Delta\\
b\hat{q}+i\Delta & 0\end{array}\right),
\label{eq:Ham_gap_mag}
\end{equation}
where now $\hat{q}=-i\frac{\partial}{\partial x}$.
\end{widetext}

\section{Derivation of the analytic solution for $E=0$}
\label{App:E0}

For the case of total momentum $K=0$, corresponding to the static exciton,
the multi-component wavefunction of relative motion satisfies
the matrix equation given by Eq.~(\ref{SchrodingerExciton}) of the main text.
Since $\phi_{AA}=\phi_{BB}$, the system of equations (\ref{SchrodingerExciton}) reduces to
\begin{equation}
\phi_{BA}=\frac{2}{\left[\varepsilon-\tilde{V}\left(x\right)\right]}\left(\hat{k}+i\Delta\right)\phi_{AA};
\label{eq:C_2_eqn}
\end{equation}
\begin{equation}
\phi_{AB}=\frac{2}{\left[\varepsilon-\tilde{V}\left(x\right)\right]}\left(\hat{k}-i\Delta\right)\phi_{AA};
\label{eq:C_3_eqn}
\end{equation}
\begin{equation}
\left(\hat{k}-i\Delta\right)\phi_{BA}+\left(\hat{k}+i\Delta\right)\phi_{AB}
=\left[\varepsilon-\tilde{V}\left(x\right)\right]\phi_{AA}.
\label{eq:Psi_AA_diff_1}
\end{equation}
Substituting Eq.~(\ref{eq:C_2_eqn}) and Eq.~(\ref{eq:C_3_eqn})
into Eq.~(\ref{eq:Psi_AA_diff_1}) yields the following second order differential equation:
\begin{equation}
\frac{d^{2}\phi_{AA}}{dx^{2}}+\frac{1}{\varepsilon-\tilde{V}(x)}\,\frac{d\tilde{V}(x)}{dx}\,\frac{d\phi_{AA}}{dx}
+\left[\left(\frac{\varepsilon-\tilde{V}(x)}{2}\right)^{2}-\Delta^{2}\right]\phi_{AA}=0,
\label{eq:Psi_AA Full}
\end{equation}
which coincides with Eq.~(\ref{PsiAA}) of the main text.

Let us now consider the case of
\[
\tilde{V}\left(x\right)=-\frac{\alpha}{\cosh\left(\beta x\right)}\]
and $\varepsilon=0$. Making the change of variable $z=\beta x$ transforms Eq.~(\ref{eq:Psi_AA Full})
to
\begin{equation}
\frac{d^{2}\phi_{AA}}{dz^{2}}+\tanh\left(z\right)\frac{d\phi_{AA}}{dz}
+\left[-\frac{1}{4}\omega^{2}\tanh^{2}\left(z\right)
+\frac{1}{4}\omega^{2}-\tilde{\Delta}^{2}\right]\phi_{AA}=0,
\label{eq:Psi_AA_Full_a}
\end{equation}
where $\omega=\alpha/\beta$ and $\tilde{\Delta}=\Delta/\beta$. The
change of variable $\chi=\tanh\left(z\right)$ allows Eq.~(\ref{eq:Psi_AA_Full_a})
to be expressed as
\begin{equation}
\left(\chi^{2}-1\right)^{2}\frac{\partial^{2}\phi_{AA}}{\partial\chi^{2}}+
\chi\left(\chi^{2}-1\right)\frac{\partial\phi_{AA}}{\partial\chi}+\left(c\chi^{2}+e\right)\phi_{AA}=0,
\label{eq:KamkeEq1}
\end{equation}
where $c=-\frac{1}{4}\omega^{2}$ and $e=\frac{1}{4}\omega^{2}-\Delta^{2}$.
Eq.~(\ref{eq:KamkeEq1}) is of a known form and has the solutions \cite{Kamke_Book}
\begin{equation}
\phi_{AA}=A_{1}\left(\chi+1\right)^{p}\left(\chi-1\right)^{q}\eta\left[\frac{1}{2}\left(\chi+1\right)\right],
\label{eq:KamkeSolution}
\end{equation}
where $A_{1}$ is a constant, the function $\eta\left[\frac{1}{2}\left(\chi+1\right)\right]$
is to be found and $p$ and $q$ are found from the following conditions:
\begin{eqnarray}
4q\left(q-1\right)+2q+c+e=0;
\nonumber\\
\qquad\left(p-q\right)\left[2\left(p+q\right)-1\right]=0.
\label{eq:PQ_Conditions}
\end{eqnarray}
Eqs.~(\ref{eq:KamkeEq1}, \ref{eq:KamkeSolution}, \ref{eq:PQ_Conditions}) yield
\begin{equation}
\left(\chi^{2}-1\right)\frac{\mathrm{d}^{2}\eta}{\mathrm{d}\chi^{2}}
+\left[\left(2p+2q+1\right)\chi-2\left(p-q\right)\right]
\frac{\mathrm{d}\eta}{\mathrm{d}\chi}
+\left[\left(p+q\right)^{2}+c\right]\eta=0.
\label{eq:before_hypergeom}
\end{equation}
Performing a change of variable $\left(\chi+1\right)/2\rightarrow\kappa$
reduces Eq.~(\ref{eq:before_hypergeom}) to the hypergeometric equation
\[\kappa\left(\kappa-1\right)\frac{\mathrm{d}^{2}\eta}{\mathrm{d}\kappa^{2}}+
\left[\left(a_{1}+a_{2}+1\right)\kappa-a_{3}\right]\frac{\mathrm{d}\eta}{\mathrm{d}\kappa}+a_{1}a_{2}\eta=0,\]
where $a_{1}=p+q\pm\sqrt{-c}$, $a_{2}=p+q\mp\sqrt{-c}$ and $a_{3}=2p+\frac{1}{2}$.
Thus, the form of $\eta$ is
\begin{equation}
\eta=\mathrm{_{2}F_{1}}\left(p+q-\sqrt{-c},\,p+q+\sqrt{-c};\,2p+\frac{1}{2};\,\kappa\right).
\end{equation}
Hence the wavefunction $\phi_{AA}$ is given by
\begin{equation}
\phi_{AA}=A_{1}\left(1+\chi\right)^{p}\left(1-\chi\right)^{q}
\mathrm{_{2}F_{1}}\left(p+q-\frac{\omega}{2},\, p+q+\frac{\omega}{2};\,2p+\frac{1}{2};\,\frac{1+\chi}{2}\right).
\label{eq:hypergeom_tobe}
\end{equation}
From Eq.~(\ref{eq:PQ_Conditions}) $q$ is found to be
\begin{equation}
q=\frac{1\pm\sqrt{1+4\tilde{\Delta}^{2}}}{4},
\end{equation}
and $p$ can take the values
\begin{equation}
p=q\end{equation}
or
\begin{equation}
p=\frac{1}{2}-q.\end{equation}
Let us first consider the case of $p=q$. In this instance Eq.~(\ref{eq:hypergeom_tobe})
becomes
\begin{equation}
\phi_{AA}=A_{1}\left[1-\chi^2\right]^{q}
\mathrm{_{2}F_{1}}\left(2q-\frac{\omega}{2},\,2q+\frac{\omega}{2};\,2q+\frac{1}{2};\,
\frac{1+\chi}{2}\right).
\label{eq:Final_form}
\end{equation}
For the function $\phi_{AA}$ to vanish as $z\rightarrow\infty$ we
require
\begin{equation}
q=\frac{1\pm\sqrt{1+4\tilde{\Delta}^{2}}}{4}>0
\end{equation}
and that the hypergeometric series is terminated. This can be satisfied
if we take the positive root of $q$ and restrict $\omega$ such that
$2q-\frac{\omega}{2}=-n$, where $n$ is a positive integer, thus we
arrive at the condition that
\begin{equation}
\omega=1+2n+\sqrt{1+4\tilde{\Delta}^{2}}.
\end{equation}
The functions $\phi_{BA}$ and $\phi_{AB}$, which can be found from Eq.~(\ref{eq:C_2_eqn})
and Eq.~(\ref{eq:C_3_eqn}), must also vanish as $z\rightarrow\infty$.
For simplicity let us analyze the linear combinations $\Psi_{+}=\phi_{BA}+\phi_{AB}$
and $\Psi_{-}=\phi_{BA}-\phi_{AB}$:
\begin{equation}
\Psi_{+}=i\frac{4\beta}{V\left(z\right)}\frac{\partial\phi_{AA}}{\partial z}=-i\frac{4}{\omega}\cosh\left(z\right)\frac{\partial\phi_{AA}}{\partial z};
\label{eq:symm_1}
\end{equation}
\begin{equation}
\Psi_{-}=-i\frac{4\Delta}{V\left(z\right)}\phi_{AA}=i\frac{4\tilde{\Delta}}
{\omega}\cosh\left(z\right)\phi_{AA}.
\label{eq:symm_2}
\end{equation}
For the functions satisfying Eqs.~(\ref{eq:symm_1}) and (\ref{eq:symm_2})
to vanish in the limit of $z\rightarrow\infty$, we require
\begin{equation}
q=\frac{1+\sqrt{1+4\tilde{\Delta}^{2}}}{4}>\frac{1}{2}.
\end{equation}
This is automatically satisfied for any finite $\tilde{\Delta}$.
Therefore, providing $\phi_{AA}$ vanishes as $z\rightarrow\infty$,
$\phi_{BA}$ and $\phi_{AB}$ will also vanish.

Let us now consider the case of $p+q=\frac{1}{2}$, in this instance
Eq.~(\ref{eq:hypergeom_tobe}) becomes
\begin{equation}
\phi_{AA}=\tilde{A}_{1}\left(1-\chi^{2}\right)^{q}
\left(1+\chi\right)^{\frac{1}{2}-2q}
\mathrm{_{2}F_{1}}\left(\frac{1}{2}-\frac{\omega}{2},\,\frac{1}{2}
+\frac{\omega}{2};\,\frac{3}{2}-2q;\,\frac{1+\chi}{2}\right).
\label{eq:hyper_p_not_q}
\end{equation}
Using the identity \cite{Abramowitz_Stegun}
\begin{equation}
z^{1-c}\mathrm{_{2}F_{1}}\left(a+1-c,\, b+1-c;\,2-c;\, z\right)=\mathrm{_{2}F_{1}}\left(a,\, b;\, c;\, z\right),
\end{equation}
where $a=2q-\frac{\omega}{2}$, $b=2q+\frac{\omega}{2}$, $c=2q+\frac{1}{2}$
and $z=\left(1+\chi\right)/2$, Eq.~(\ref{eq:hyper_p_not_q}) becomes
\begin{equation}
\phi_{AA}=\tilde{A}_{2}\left[1-\chi^{2}\right]^{q}
\mathrm{_{2}F_{1}}\left(2q-\frac{\omega}{2},\,2q+\frac{\omega}{2};\,2q+\frac{1}{2};
\,\frac{1+\chi}{2}\right),
\end{equation}
where $\tilde{A}_{2}$ is a constant. This is of the same form as
Eq.~(\ref{eq:Final_form}).

\section{Binding energy in the limit of weak electron-hole attraction}
\label{App:weak}
Let us consider the limit of very weak electron-hole attraction: $|\tilde{V}(x)| \ll \Delta $ for any
inter-particle separation. Substituting $\varepsilon=2 \Delta - \varepsilon_{b}$ into Eq.~(\ref{PsiAA})
and retaining only the first order terms in $\tilde{V}/\Delta$ and $\varepsilon_{b}/\Delta$ transforms
Eq.~(\ref{PsiAA}) into
\begin{equation}
-\frac{1}{\Delta}\frac{d^{2}\phi_{AA}}{dx^{2}}+\tilde{V}(x)\,\phi_{AA}
=-\varepsilon_{b} \phi_{AA}.
\label{weakSchr}
\end{equation}
Eq.~(\ref{weakSchr}) is of the form of the non-relativistic one-dimensional Schr\"{o}dinger equation,
where the parameter $\Delta$, plays the role of the effective mass. For a particle of mass $m$ in a
weak one-dimensional potential $U(x)$, the textbook\cite{Landau} treatment of the potential as a
perturbation yields the following result for the binding energy:
$|E|=(m/2\hbar^2)\left[\int_{-\infty}^{\infty}U(x) dx \right]^2$, which can be reformulated for
Eq.~(\ref{weakSchr}) as
\begin{equation}
\varepsilon_{b}=\frac{\Delta}{4}\left[\int_{-\infty}^{\infty}\tilde{V}(x) dx \right]^2.
\label{universalGeneral}
\end{equation}
For $\tilde{V}(x)=-\alpha/\cosh(\beta x)$,
Eq.~(\ref{universalGeneral}) yields
\begin{equation}
\varepsilon_{b}=\frac{\pi^2}{4} \Delta\, \left(\alpha/\beta\right)^2.
\label{universalHyperSecant}
\end{equation}
Thus, we have shown that for a weak hyperbolic secant potential, when $\alpha \ll \Delta$, the binding energy
scales with the band gap and has the universal quadratic dependence on $\alpha/\beta$.
Notably, Eq.~(\ref{universalHyperSecant}) is also true for a weak interaction potential
given by a Lorentzian, $\tilde{V}(x)=\alpha/\left(1+\beta^2 x^2\right)$.

\section{Exciton with attractive delta-function potential}
\label{App:deltafunction}

The problem of calculating the energy of an exciton in a carbon nanotube can be solved exactly in the case when interaction potential between the electron and hole is taken to be a delta-function, $\tilde{V}(x)=-U_0\delta(x)$, where the strength of the potential $U_0$ is positive and can be estimated as a product of the strength of the realistic potential and its width.

Let us consider an electron-hole pair in a narrow-gap 1D carbon nanotube with a band gap energy of $4|\Delta|$.
Measuring all quantities in units of $\hbar v_F$ as above, one can write the Hamiltonian as
\begin{equation}
\widehat{H}=\sqrt{\Delta^2+\widehat{q}_e^2}+\sqrt{\Delta^2+\widehat{q}_h^2}+\tilde{V}(|x_e-x_h|),
\label{Hinitial}
\end{equation}
where
\begin{equation}
\widehat{q}_{e,h}=-i\frac{\partial}{\partial x_{e,h}}.
\end{equation}
Here we have retained only the eigenstates of the general graphene-type Hamiltonian with positive energies.
The Schrodinger equation for the problem we consider thus reads
\begin{equation}
\left[\sqrt{\Delta^2+\widehat{q}_e^2}+\sqrt{\Delta^2+\widehat{q}_h^2}+\tilde{V}(|x_e-x_h|)\right] 
\Psi(x_e,x_h)=\varepsilon \Psi(x_e,x_h).
\label{ShrodingerDelta}
\end{equation}
Introducing new variables corresponding to the center of mass
and relative motion in the manner explained in the text of the article and putting the wavevector of the center of mass motion equal to zero, one gets the following expression
\begin{equation}
2\sqrt{\Delta^2-\frac{d^2}{dx^2}}\psi(x)=\left[\varepsilon-\tilde{V}(x)\right]\psi(x),
\label{SchrodingerDelta1}
\end{equation}
where $\psi(x)$ is a wavefunction of relative motion. The above expression is a 1D Schr\"{o}dinger
equation for a particle with a complicated dispersion placed in an
external potential $\widetilde{V}(x)$, which can be solved for the case when $\tilde{V}(x)=-U_0\delta(x)$. Indeed, in this case the solution of the Schr\"{o}dinger equation corresponding to a bound state reads
\begin{eqnarray}
\psi(x)=Ae^{\kappa(\epsilon)x},~x<0, \label{PsiDelta1}\\
\psi(x)=Ae^{-\kappa(\epsilon)x},~x>0, \label{PsiDelta2}
\end{eqnarray}
where $\varepsilon=2\sqrt{\Delta^2+\kappa^2}$. It is well known that the zero-range delta
potential is equivalent to the introduction of the specific boundary condition for the derivative
of the solution at  $x=0$ (the function itself should be continuous, $\psi(+0)=\psi(-0)$).
To obtain a condition for the derivative, let us integrate the Schr\"{o}dinger equation
\begin{equation}
\left[\widehat{H}_0\left(\frac{d}{dx}\right)-\varepsilon\right]\psi(x)=U_0\delta(x)\psi(x),
\end{equation}
where $\widehat{H}_0\left(\frac{d}{dx}\right)=2\sqrt{\Delta^2+\frac{d^2}{dx^2}}$ across
the interval $[-0;+0]$; this yields
\begin{equation}
\int{\widehat{H}_0\left(\frac{d}{dx}\right)\psi(x)dx}|_{x=+0}-\int{\widehat{H}_0\left(\frac{d}{dx}\right)
\psi(x)dx}|_{x=-0}=U_0\psi(0).
\end{equation}
Using this relation and Eqs.~(\ref{PsiDelta1}), (\ref{PsiDelta2})
and (\ref{SchrodingerDelta1}) one gets the following expression for
determining the parameter $\kappa$
\begin{equation}
H_0(-\kappa)+H_0(\kappa)-2H_0(0)=-\kappa U_0.
\end{equation}
Thus,
\begin{equation}
\sqrt{\Delta^2-\kappa^2}=\Delta-\frac{\kappa U_0}{4},
\label{kappa_eq_K0}
\end{equation}
which yields
\begin{equation}
\kappa=\frac{8U_0\Delta}{16+U_0^2},
\label{kappa_K0}
\end{equation}
where the energy is given by
\begin{equation}
\varepsilon=2\sqrt{\Delta^2-\kappa^2}.
\label{energy_K0}
\end{equation}
Note, that this result is valid only for the case of $U_0<4$, as for $U_0>4$, the right hand side of
Eq.~(\ref{kappa_eq_K0}) is negative.
The binding energy of the exciton can be determined as
\begin{equation}
\varepsilon_b=2\Delta-\epsilon=4\Delta U_0^2/(16+U_0^2),
\label{BindingEnergy}
\end{equation}
which tends to the non-relativistic result of $\varepsilon_b=U_0^2 \Delta/4$ for $U_0 \ll 1$; c.f. this result with
Eq.~(\ref{universalGeneral}). Note, that in both relativistic and non-relativistic cases the binding energy is proportional to the gap.


\end{document}